\documentclass[12pt]{article}
\usepackage{graphicx}

\title{MAGNETIC COMPOSITES: MAGNONIC EXCITATIONS vs. THREE-DIMENSIONAL STRUCTURAL PERIODICITY}
\author{M. Krawczyk\footnote{Corresponding author; e-mail: krawczyk@amu.edu.pl} $\;$ and H. Puszkarski}
\date{}

\begin{document}
%\draft

\begin{titlepage}

%\address{Surface Physics Division, Faculty of Physics, Adam Mickiewicz University,  ul. Umultowska 85, Pozna\'{n}, %61-614 Poland.}
\maketitle \thispagestyle{empty} \vspace{-1cm}
\begin{center}
Surface Physics Division, Faculty of Physics, Adam Mickiewicz University, \\ ul. Umultowska 85, 61-614 Pozna\'{n}, Poland
\end{center}

\baselineskip=25pt
\section*{ABSTRACT}
This study deals with the spin wave spectrum in magnetic macrostructure (composed of two ferromagnetic materials) showing a 3D periodicity:
spherical ferromagnetic grains disposed in the nodes of a simple cubic crystal lattice are embedded in a matrix with different ferromagnetic
properties. It is shown that the \textit{magnonic spectrum} of this composite structure exhibits frequency regions \textit{forbidden} for magnon
propagation, and the energy gaps are found to be sensitive to the exchange contrast between the constituent materials as well as to the
magnetization contrast. The widths of the respective magnonic gaps are studied as
functions of parameters characterizing the magnetic structure.\\
\\
{\bf Keywords}: magnonic crystals, spin waves, periodic composites \\
{\bf PACS numbers}: 75.50.-y; 75.30.Ds; 75.40.Gb\\

\end{titlepage}

\newpage
\baselineskip=25pt
\section*{INTRODUCTION}

Although the first study of electromagnetic wave propagation in periodic structures, written by Lord Rayleigh, was published already in 1887, it
is only in recent years that photonic composites raised suddenly an extremely interest. The research in this field was initiated by the studies
by Yablonovitch and John \cite{[1],[2]} published in 1987 and anticipating the existence of complete energy gaps in electromagnetic wave spectra
in three-dimensional periodic composites, henceforth referred to as {\em photonic crystals}. These can be used for fabricating new
optoelectronic devices in which the role of electrons, traditionally used as transport medium, would be played by photons \cite{[3],[4],[5]}.
The so-called {\em left-handed materials} (LHM), showing negative effective refractivity \cite{[6],[7]}, provide an example of periodic
materials demonstrating how much the properties of this kind of structure can differ from those of homogeneous materials. Another type of
periodic composites are structures composed of materials with different {\em elastic} properties; showing an energy gap in their {\em elastic}
wave spectrum, such composites are referred to as {\it phononic crystals} \cite{[8],[9],[10]}. Recently, attention has been focused on the
search of photonic and phononic crystals in which both the position and the width of the energy gap could be controlled by external factors,
such as applied voltage or magnetic field. Attempts have been made to create photonic crystals in which one of the component materials is a
magnetic \cite{[11],[12],[13],[14],[15],lyubchanskii03}.

A magnetic periodic composite consists of at least two magnetic materials; the information carrier in such structures are spin waves. By analogy
to photonic and phononic crystals, in which the role of information carrier is played by photons and phonons respectively, periodic magnetic
composites are referred to as {\em magnonic crystals}. Studies of 2D magnonic crystals have been reported \cite{[16],[17],[18],[19],[20],[21]},
with scattering centers in the form of "infinitely" long cylinders disposed in square lattice nodes (cylinder and matrix materials being two
different ferromagnetics), and the anticipated gaps were found indeed in the respective spin wave spectra. Further research was focused on
magnetic multilayer systems, which can be regarded as 1D magnonic crystals \cite{[22],[23],[24],[25],[26],[27],[28],[29],[30]}.

In this paper, we present numerically calculated magnonic band structures of {\em three-dimensional} magnonic crystals. Due to the complexity of
the problem, only the simplest model of 3D magnonic crystal is considered here, represented by a system of ferromagnetic spheres (which act as
scattering centers) disposed in the nodes of a simple cubic crystal lattice and embedded in a different ferromagnetic material (matrix). Both
the exchange and dipolar interactions are taken into account in our calculations based on the plane wave method and using the linear
approximation.

\section*{THEORY OF 3D MAGNONIC BAND STRUCTURE}

\begin{figure}[h]
\begin{center}
\includegraphics[width=120mm]{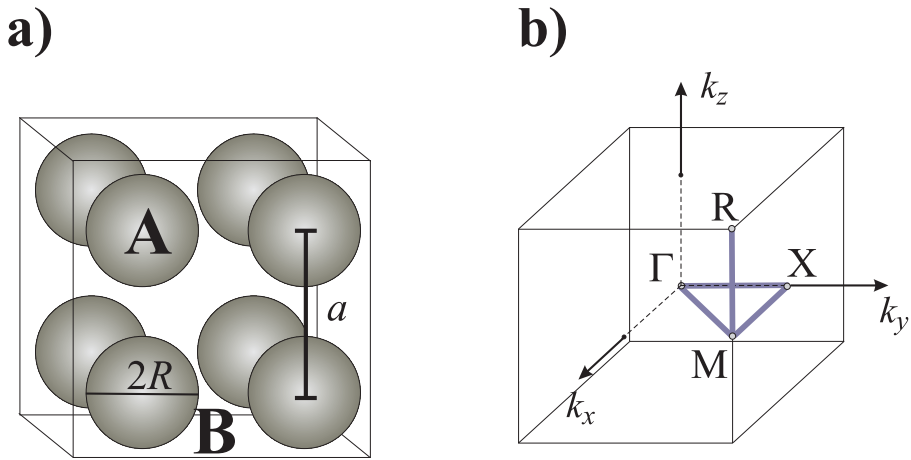}
%\vspace{2mm}
\caption{a) The 3D periodic structure studied in this paper; the structure consists of ferromagnetic spheres of material {\bf A} embedded in a
matrix of material {\bf B} (materials {\bf A} and {\bf B} have different magnetic properties); the spheres are disposed in the nodes of a {\em
sc} lattice. (b) The Brillouin zone corresponding to the considered structure, and the path $M \Gamma XMR$ (highlighted) along which the
magnonic band spectra are calculated.\label{fig1}}
\end{center} \end{figure}

Let's consider an ideal periodic structure consisting of spheres of ferromagnetic {\bf A} and embedded in a matrix of ferromagnetic {\bf B}. The
spheres are assumed to form a 3D periodical lattice of  {\em sc} type (Fig. \ref{fig1}a). A static magnetic field, $H_{0}$, is applied to the
composite along the $z$ axis, and assumed to be strong enough to saturate the magnetization of both materials. The lattice constant is denoted
by $a$; the filling fraction, $f=4/3 \pi R^{3} a^{-3}$, is defined as the volume proportion of material {\bf A} in a unit cell.

Ferromagnetics {\bf A} and {\bf B} are characterized by two
material parameters: the spontaneous magnetization ($M_{S,A}$ and
$M_{S,B}$), and the exchange constant ($A_{A}$ and $A_{B}$); both
these parameters depend on the position vector $\vec{r}=(x,y,z)$:
\begin{eqnarray}
M_{S}(\vec{r})&=&M_{S,B}+(M_{S,A}-M_{S,B})S(\vec{r}),
\nonumber \\
A(\vec{r})&=&A_{B}+(A_{A}-A_{B})S(\vec{r}),\label{eq2}
\end{eqnarray}
the value of function $S(\vec{r})$ being 1 inside a sphere, and 0
beyond.

In the classical approximation spin waves are described by the
Landau-Lifshitz (LL) equation, taking the following form in the
case of magnetic composites (with damping neglected):
\begin{eqnarray}
\frac{\partial \vec{M}(\vec{r},t)}{\partial t} = \gamma \mu_{0}
\vec{M}(\vec{r},t) \times \vec{H}_{eff}(\vec{r},t), \label{eq3}
\end{eqnarray}
where magnetization $\vec{M}(\vec{r},t)$ is a function of position
vector $\vec{r}$ and time $t$; $\vec{H}_{eff}(\vec{r},t)$ stands
for the effective magnetic field  \cite{[17],[18],[19],landau35}
acting on magnetization $\vec{M}(\vec{r},t)$:
\begin{eqnarray}
\vec{H}_{eff}(\vec{r},t)=H_{0}\hat{z}
+\vec{h}(\vec{r},t)+\frac{2}{\mu_{0}} \left(\nabla \cdot
\frac{A}{M^{2}_{S}}\nabla \right) \vec{M}(\vec{r},t); \label{eq4}
\end{eqnarray}
$\hat{z}$ is the unit vector along the $z$ axis; $\vec{h}(\vec{r},
t)$ is the dynamic magnetic field due to dipolar interactions; the
third component represents the exchange field. The magnetization
vector can be represented as the sum of its static component,
$M_{S}\hat{z}$, which is parallel to the applied field, and its
dynamic component, $\vec{m}(\vec{r},t)$, lying in plane ($x,y$):
\begin{eqnarray}
\vec{M}(\vec{r},t)=M_{S}\hat{z}+\vec{m}(\vec{r},t). \label{eq5}
\end{eqnarray}
The dynamic dipolar field, $\vec{h}$, must satisfy the
magnetostatic Maxwell equations:
\begin{eqnarray}
\nabla \times \vec{h}(\vec{r})=0, \nonumber \\
\nabla \cdot \left(\vec{h}(\vec{r})+ \vec{m}(\vec{r})\right)=0.
\label{eq6}
\end{eqnarray}
In a magnonic crystal, the position-dependent coefficients in (\ref{eq4}), {\em i.e.} $M_{S}$ and $A$, are periodic functions of the position
vector, which allows us to use, in the procedure of solving the LL equation defined in (\ref{eq3}), the plane wave method, described in detail
in our earlier papers \cite{[17],[18]} (dealing with {\em two-dimensional} magnonic crystals). Following this scheme, we proceed to
Fourier-expanding all the periodic functions of the position vector, {\em i.e.} the spontaneous magnetization, $M_{S}$, and parameter $Q$
defined as follows:
\begin{eqnarray}
Q=\frac{2A}{\mu_{0}M^{2}_{S}H_{0}}. \label{eq7}
\end{eqnarray}

The dynamic components of the magnetization can be expressed as the product of the periodic envelope function and the Bloch factor
$exp(i\vec{k}\vec{r}$) ($\vec{k}$ denoting a 3D wave vector); the envelope function can be transformed into the reciprocal space as well.
Including all the expansions into (\ref{eq3}) and (\ref{eq6}) leads to the following infinite system of linear equations for Fourier
coefficients of the dynamic magnetization components, $\vec{m}_{x \vec{k}}(\vec{G})$  and $\vec{m}_{y \vec{k}}(\vec{G})$:
%\begin{widetext}
\begin{eqnarray}
 i\Omega m_{x \vec{k}}(\vec{G})= m_{y \vec{k}}(\vec{G})
+\sum_{\vec{G}'} \frac{(k_{y} + G'_{y})(k_{x}+G'_{x})
m_{x\vec{k}}(\vec{G}') + (k_{y} + G'_{y})^{2}m_{y
\vec{k}}(\vec{G}')}{H_{0}| \vec{k}+
 \vec{G}' |^{2}} M_{S}(\vec{G}- \vec{G}')  \nonumber
\\ +\sum_{\vec{G}'} \sum_{\vec{G}''} [ (\vec{k} + \vec{G}') \cdot (\vec{k}+\vec{G}'')
- (\vec{G}-\vec{G}'') \cdot (\vec{G} - \vec{G}' ) ]
  M_{S}(\vec{G}-\vec{G}'') Q(\vec{G}'' - \vec{G}') m_{y \vec{k}} (\vec{G}'),\nonumber
\\
 i \Omega m_{y \vec{k}}(\vec{G})= -m_{x \vec{k}} (\vec{G})-\sum_{\vec{G}'}
\frac{(k_{y} + G'_{y})(k_{x}+G'_{x}) m_{y
\vec{k}}(\vec{G}')+(k_{x} + G'_{x})^{2}m_{x
\vec{k}}(\vec{G}')}{H_{0}| \vec{k} + \vec{G}' |^{2}}
 M_{S}(\vec{G} -\vec{G}') \nonumber \\
- \sum_{\vec{G}'} \sum_{\vec{G}''} [ (\vec{k} + \vec{G}') \cdot (\vec{k}+\vec{G}'') - (\vec{G}-\vec{G}'') \cdot (\vec{G} - \vec{G}' ) ]
M_{S}(\vec{G}-\vec{G}'') Q(\vec{G}'' - \vec{G}') m_{x \vec{k}} (\vec{G}'),\nonumber \\
 \label{eq8}
\end{eqnarray}
%\end{widetext}
$k_{x},k_{y}$ and $G_{x},G_{y}$ denoting the Cartesian components of the wave vector $\vec{k}$ and a reciprocal lattice vector $\vec{G}$,
respectively; a new quantity introduced in (\ref{eq8}) is $\Omega$, henceforth referred to as {\em reduced frequency}:
\begin{eqnarray}
\Omega=\frac{\omega}{|\gamma| \mu_{0} H_{0}}. \label{eq9}
\end{eqnarray}

The Fourier coefficients of spontaneous magnetization $M_{S}$ and
parameter $Q$ are calculated from the inverse Fourier
transformation; in the case of spheres the resulting formulae read
for $M_{S}$ as follows:
\begin{eqnarray*}
M_{S}(\vec{G}) = \left\{ \begin{array}{l}
     M_{S,A}f + M_{S,B}(1-f),\;
     \mbox{for  $\vec{G}=0$} \\
f  (M_{S,A}-M_{S,B}) {\displaystyle \frac{3\left[\sin(GR)-(GR)\cos(GR)\right]}{(GR)^{3}}},
\mbox{ \hspace{0.5cm}for $\;\vec{G} \neq 0$}%
    \end{array}
    \right.%
     \label{aa1}
\end{eqnarray*}
and similarly for $Q$; $R$ is the sphere radius (Fig.
\ref{fig1}a).

Obviously, the numerical calculations performed on the basis of
(\ref{eq8}) involve a finite number of reciprocal lattice vectors
$\vec{G}$ in the Fourier expansions; however, we have made sure
that the number used is large enough to ensure good convergence of
the numerical results. As indicated by an analysis performed, a
satisfactory convergence is obtained already with 343 reciprocal
lattice vectors used.

\begin{figure}[h]
\begin{center}
\includegraphics[width=70mm]{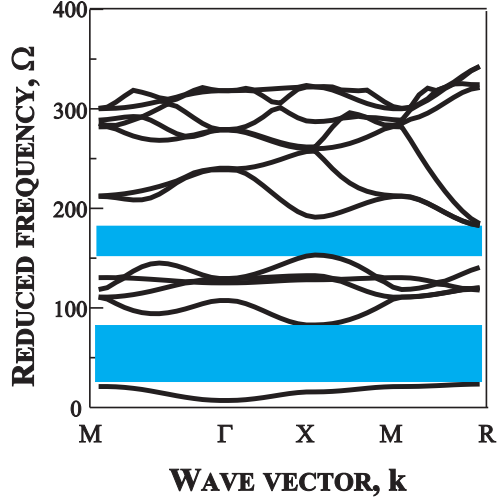}
%\vspace{2mm}
\caption{The magnonic band structure found numerically for the
$sc$ lattice-based composite (iron spheres embedded in a YIG
matrix). The spin wave energy branches have been calculated along
the path (in the first Brillouin zone) shown in Fig. \ref{fig1}b
for the filling fraction $f$=0.2. The lattice constant value is assumed to be $a=100\AA$,
and the applied magnetic field value is $\mu_{0}H_{0}=0.1$T.
\label{fig2}}
\end{center} \end{figure}

\section*{NUMERICAL RESULTS}

The 3D magnonic crystal studied is a magnetic composite consisting of ferromagnetic spheres (of material {\bf A}) disposed in the nodes of a
{\em sc} lattice and embedded in a different magnetic material ({\bf B}), referred to as matrix (Fig. \ref{fig1}). The corresponding magnonic
band structure will be calculated along path $M=\pi/a(1,1,0) \rightarrow \Gamma= \pi/a(0,0,0) \rightarrow X=\pi/a(0,1,0) \rightarrow
M=\pi/a(1,1,0) \rightarrow R=\pi/a(1,1,1)$ in the nonreducible part of the first Brillouin zone (see Fig. \ref{fig1}). Iron  (Fe) and yttrium
iron garnet (YIG) are chosen as component materials {\bf A} (spheres) and {\bf B} (matrix), respectively, in the studied example. As established
in our earlier studies \cite{[17],[18]} (in the case of 2D magnonic crystals) such composition, involving a substantial contrast of the magnetic
parameters between the component materials ($M_{S,Fe}=1.752\; 10^{6} Am^{-1}$, $M_{S,YIG}=0.194 \; 10^{6}Am^{-1}$, $A_{Fe}=2.1 \;
10^{-11}Jm^{-1}$, $A_{YIG}=0.4 \; 10^{-11}Jm^{-1}$), leads to opening large energy gaps in the spin wave spectrum. Also in the spin wave
spectrum obtained for the {\em sc} lattice-based 3D magnonic crystal considered here, two wide energy gaps are present, one between bands 1 and
2, the other between bands 5 and 6, as shown in Fig. \ref{fig2}. The existence of the gap signifies that spin waves having frequency values
within the gap cannot propagate through the composite. The following parameter values have been assumed in our calculations: crystal lattice
constant $a$=100$\AA$, applied static magnetic field $\mu_{0}H_{0}=0.1$T, and filling fraction $f$=0.2.

\begin{figure}[h]
\begin{center}
\includegraphics[width=130mm]{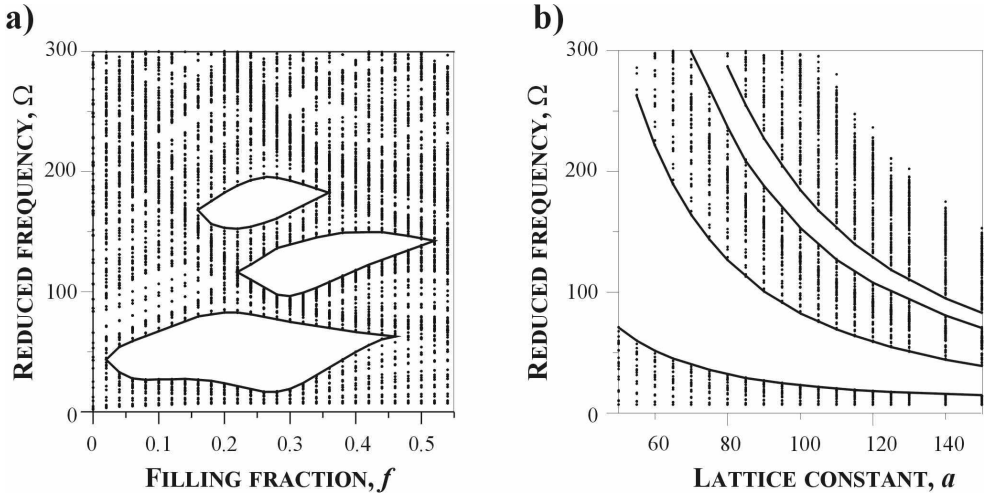}
%\vspace{2mm}
\caption{(a) The magnonic energy branches in the Fe/YIG simple cubic magnonic crystal plotted \textit{versus} the filling fraction. The lattice
constant value is assumed to be $a=100\AA$, and the applied magnetic field value is $\mu_{0}H_{0}=0.1$T. (b) Magnonic branches plotted versus
the lattice constant $a$ of the same {\em sc} structure; the filling fraction is fixed at $f = 0.2$. A frequency gap (the white region) moves
down with increasing $a$. \label{fig3}}
\end{center} \end{figure}

Let's examine the effect of the filling fraction on the width of the spectral gaps. Figure \ref{fig3}a shows the magnonic bands plotted against
the filling fraction in the {\em sc} lattice-based structure ($Fe$ spheres embedded in YIG). The plot was obtained through projecting the
magnonic band structure calculated along $M\Gamma XMR$ path (for a fixed filling fraction value) on the reduced frequency ($\Omega$) axis;
afterwards the procedure of projection was performed repeatedly for consecutive filling fraction values. The maximum width of the first gap is
found to occur at $f$=0.27, reaching value $\Delta \Omega$=63.02. Unlike in 1D and 2D magnonic crystals \cite{[17],[18],[22]}, no oscillatory
variations of the energy gaps with the filling fraction are observed in the considered 3D composite. Figure \ref{fig3}b shows the computed
magnonic band structure plotted against the lattice constant $a$ for $f=0.2$. The gap center ($\Omega_{0}$) is found to descend, and the gap
itself $\Delta \Omega$ (the white region) to narrow as the lattice constant increases, in such a way that the reduced gap width $\Delta \Omega /
\Omega_{0}$ remains almost constant.

\begin{figure}[h]
\begin{center}
\includegraphics[width=140mm]{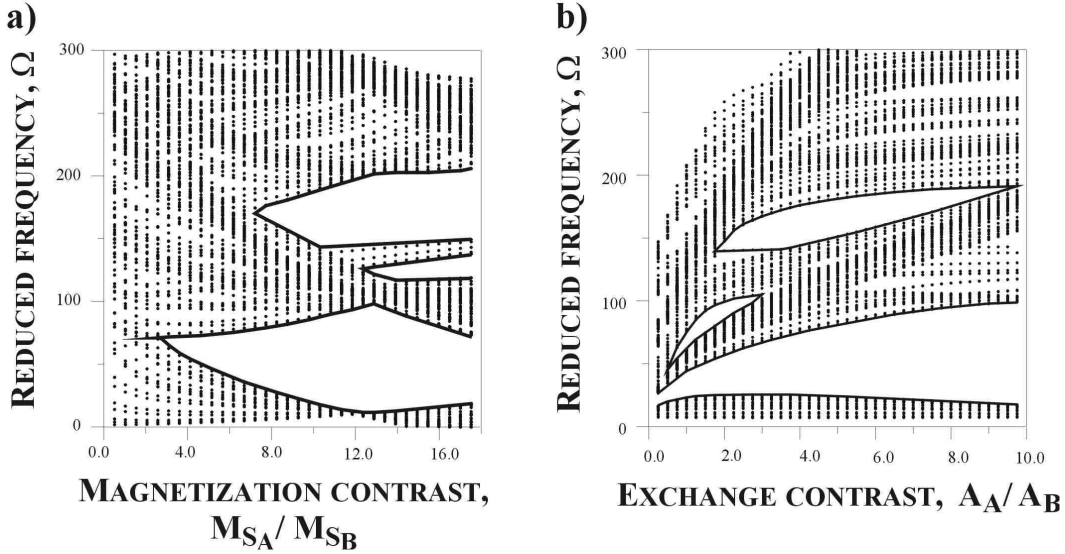}

%\vspace{2mm}

\caption{(a) Magnonic spectra plotted \textit{versus} the magnetization contrast for a $sc$ lattice-based structure. The exchange constant value
in the spheres is assumed to be $A_{A}=2.1\; 10^{-11}Jm^{-1}$. (b) Magnonic branches plotted versus the exchange contrast; the spontaneous
magnetization in the spheres is assumed to be $M_{S,A}=2.1\; 10^{6}Am^{-1}$. The assumed spontaneous magnetization and exchange constant values
of the matrix are: $M_{S,B}=0.194\; 10^{6}Am^{-1}$ and $A_{B}=0.4\; 10^{-11}Jm^{-1}$; the lattice constant $a=100\AA$, the filling fraction
$f=0.2$, and the applied field $\mu_{0}H_{0}=0.1T$. \label{fig4}}
\end{center} \end{figure}

Let's proceed now to the role played by magnetic parameters characterizing component materials. Figure \ref{fig4}a show spin wave energy
branches plotted against the contrast between the spontaneous magnetization values in materials {\bf A} (spheres) and {\bf B} (matrix); this
contrast, defined as ratio $M_{S,A}/M_{S,B}$, will be henceforth referred to as {\em magnetization contrast}. The computations have been
performed for a fictitious matrix material whose spontaneous magnetization and exchange constant values, $M_{S,B}$ and $A_{B}$, respectively,
are fixed at values close to those in YIG; the exchange constant value in the sphere material is assumed to be $A_{A}=2.1\; 10^{-11} Jm^{-1}$.
The other parameters are fixed as well: the lattice constant value is $a$=100$\AA$, the filling fraction $f$=0.2, and the applied magnetic field
$\mu_{0}H_{0}=0.1$T.  It is seen that the gap width is maximal for the magnetization contrast value 13, which is not far from the magnetization
contrast between iron and YIG ($M_{S,Fe} / M_{S,YIG}$ = 10.82); note also that the required value of magnetization contrast necessary for the
energy gap to open is greater than 2.

We shall now examine the role of the contrast between the exchange constant values in the component materials. The results of the respective
computations are depicted in Fig. \ref{fig4}b, showing magnonic branches {\em versus} the {\em exchange contrast}, defined as the ratio
($A_{A}/A_{B}$) of exchange constant values in materials {\bf A} (spheres) and {\bf B} (matrix); the only variable in the computations was the
exchange constant in material {\bf A}. All other structural and material parameters are fixed at the following values: $M_{S,A}=1.752\;
10^{6}Am^{-1}$, $M_{S,B}=0.194\; 10^{6}Am^{-1}$, $A_{B}=0.4\; 10^{-11}Jm^{-1}$, $a=100\AA$, $\mu_{0}H_{0}=0.1$T. From the Fig. \ref{fig4}b one
can infer that the exchange contrast is not an indispensable factor for the opening of complete energy gaps in 3D magnonic crystals.

\section*{CONCLUSIONS}

We demonstrated that the 3D \textit{magnonic crystal} with a {\it sc} lattice-based structure exhibits complete energy gaps, whose appearance is
due to \textit{two magnetic contrasts}: (1) the \textit{exchange} contrast  and (2) the {\it magnetization} contrast. Gaps are found to exist
only when the spontaneous magnetization contrast is greater than 2. However, the gap width can be also controlled through adjusting the
respective structural factors: the filling fraction value and the composite lattice constant.

\subsection*{ACKNOWLEDGEMENT}

The present work was supported by the Polish State Committee for
Scientific Research through projects KBN - 2P03B 120 23 and
PBZ-KBN-044/P03-2001.

%%%%%%%%%%%%%%%%%%%%%%%%%%%%%%%%%%%%%%%%%%%%%%%%%%%%%%%%%%%%%%%%%%%%%%%%%%%%
%%%%%%%%%%%%%%%%%%%%%%%%%%%%%%%%%%%%%%%%%%%%%%%%%%%%%%%%%%%%%%%%%%%%%%%%%%%%

\end{document}